\documentclass[12pt,cites,graphicx]{article}
\usepackage{epsfig}
\title{Halide Adsorption on Single-crystal Silver Substrates:  Dynamic Simulations\\
and {\it ab-initio} Density-functional Theory}
\author{S.~J.\ Mitchell, Sanwu Wang, and P.~A.\ Rikvold\\[3mm]
Center for Materials Research and Technology,\\[1mm]
School of Computational Science and Information Technology,\\[1mm]
and Department of Physics,\\[1mm]
Florida State University, Tallahassee, Florida, 32316-4120, USA\\[1mm]
}
\begin{document}
\maketitle
\renewcommand{\thefootnote}{\fnsymbol{footnote}}

\noindent
We investigate the static and dynamic behaviors of a Br adlayer electrochemically deposited onto single-crystal Ag(100)
using an off-lattice model of the adlayer.
Unlike previous studies using a lattice-gas model,
the off-lattice model allows adparticles to be located at any position
within a two-dimensional approximation to the substrate.
Interactions with the substrate are approximated by a corrugation potential.
Using Density Functional Theory (DFT) to calculate surface binding energies,
a sinusoidal approximation to the corrugation potential is constructed.
A variety of techniques, including Monte Carlo and Langevin simulations,
are used to study the behavior of the adlayer.
The lateral root-mean-square (rms) deviation of the adparticles from the binding sites
is presented along with equilibrium coverage isotherms,
and the thermally activated Arrhenius barrier-hopping model used in previous dynamic Monte Carlo
simulations is tested.

\section{Introduction}
\label{intro}

Electrochemical adsorption processes in monolayer or submonolayer systems
contain essentially four types of microscopic processes:
lateral surface diffusion, adsorption/desorption, molecular reorientation, and reaction or charge transfer.
Often, the free-energy landscape for these systems has a very large number of local minima,
each of which corresponds to a well-localized state in the full phase space.
For example, adparticles are often well localized on the crystal substrate,
spending the majority of their time near specific adsorption sites.
In these systems, the time scale for transitions between neighboring free-energy
minima is associated with hopping across a free-energy barrier
that corresponds to a saddle point in the free-energy landscape.
Dynamic Monte Carlo simulations exploit these well-localized states
and simulate the time evolution of the system as a series of thermally activated
stochastic transitions over the free-energy saddle points.

In this paper we investigate the validity of the localized-state approximation
and the commonly used Arrhenius form for the stochastic transition rate between
local free-energy minima by constructing an off-lattice model for the adsorption
of Br onto the (100) surface of an Ag single crystal.
This system has the advantage of simplicity,
in that the only microscopic processes are lateral surface diffusion and adsorption/desorption.
In this system, the only reaction or charge-transfer process is associated with adsorption/desorption,
and reorientation is not important for atomic adsorbates.

The equilibrium properties of Br adsorbed onto Ag(100) have been extensively
studied both by experiments \cite{VALET78,OCKO97,ENDO99}
and by theory \cite{WANG97,KOPE98,MITC00A,MITC00C}.
The time evolution of a lattice-gas approximation to the adsorbate dynamics
has also been extensively studied by dynamic Monte Carlo simulations \cite{MITC00A,MITC00C,MITC00B}.
The lattice-gas approximation used in Refs.~\cite{WANG97,KOPE98,MITC00A,MITC00C,MITC00B}
explicitly assumes well-localized surface states,
which is not an unreasonable assumption since this system is known to display only commensurate phases \cite{OCKO97}.
The present paper is primarily concerned with testing the well-localized state
and Arrhenius transition-rate assumptions of Dynamic Monte Carlo
for surface diffusion processes.
However, the computational techniques used here can also be applied
to adsorption/desorption and molecular reorientation barrier-hopping processes.
The off-lattice model developed here,
which does not assume well-localized surface adsorption sites,
can also be used for systems which do not have strongly localized adsorption sites,
such as the incommensurate phases observed for Br/Au(100) \cite{WAND96}.
Some preliminary results of the work presented here were included in Ref.~\cite{MITCH01A}.

The remainder of this paper is organized as follows.
In Sec.~\ref{sec:DFT}, Density Functional Theory (DFT) pseudopotential calculations
are used to calculate the surface binding energies and deduce an approximate corrugation potential for Br on Ag(100).
In Sec.~\ref{sec:off}, an off-lattice model is constructed using the corrugation potential,
and the well-localized surface-state assumption is tested by equilibrium
simulation of this off-lattice model.
In Sec.~\ref{sec:arrhenius},
the thermally activated stochastic Arrhenius transition-rate
assumption is tested for diffusion processes by simulating single-particle
trajectories within the corrugation potential by Langevin simulations.
In Sec.~\ref{sec:conc}, our conclusions are presented,
and in the Appendix, the grand-canonical off-lattice Monte Carlo method is explained in detail.

\section{Corrugation Potential}
\label{sec:DFT}

{\it In situ} Surface X-ray Scattering (SXS) \cite{OCKO97} and X-ray Absorption Fine Structure (XAFS) \cite{ENDO99}
measurements of Br deposited on Ag(100) in solution, as well as Low Energy Electron Diffraction (LEED)
measurements \cite{KLEINHERBERS89} in vacuum confirm that Br adsorbs at the four-fold hollow sites of the substrate,
indicating that the corrugation potential has its minima at these points, both in vacuum and in solution.
However, knowledge of the binding site alone is not enough to determine the symmetry
or amplitude of the approximate corrugation potential,
and therefore {\it ab initio} Density Functional Theory (DFT) pseudopotential calculations
were used to determine the parameters of the corrugation potential.

We point out that one must be careful in selecting
surface models for {\it ab initio} calculations, as the binding
energy is very sensitive to the way the surface is modeled (cluster or slab models).
This point is explained in further detail in Ref.~\cite{SWANG01}. We
emphasize that calculations with small metal clusters should be used with
caution since the electronic structures of a small metal cluster and a 
surface of the same metal are usually very different.
Small cluster models are useful in providing detailed information
about the local bonding. However, convergence of the binding energy
with the cluster size is usually difficult to achieve \cite{WHITTEN96},
and cluster calculations can even lead to predictions in direct conflict with experimental
results, such as for the binding site of Br on Au(100) \cite{IGNA98}. In
order to simulate metal surfaces accurately large clusters or
extended surface models (e.g., slab models) are therefore needed. In contrast to metals, 
small clusters are usually quite good models for
semiconductors since the electronic states and the
chemical bonds are much more localized in this case \cite{SWANG02}.

In our DFT calculations, the Ag(100) surface was modeled by repeated slabs
with seven metal layers of nine Ag atoms separated by a vacuum region
equivalent to seven metal layers, corresponding to a $3\times3$ surface
cell. Br was placed symmetrically on both sides of the slab. Total
energies for the hollow, bridge, and on-top configurations, as well as
for configurations with Br at the midpoints between the hollow and bridge, the
hollow and on-top, and the bridge and on-top sites were calculated. All
the metal atoms were located at their bulk positions, with the equilibrium
bulk lattice constant of 4.17~{\AA} determined by our calculations, 
as compared to the experimental value of 4.09~{\AA} \cite{CRSH90}. Test
calculations showed that relaxation of the Ag positions had a negligible
effect on the total-energy differences between the hollow and bridge
configurations, and we therefore relaxed only the Br position in the
direction perpendicular to the surface.

Our {\it ab initio} total-energy calculations were performed within
density functional theory with the Vienna {\it ab-initio} simulation 
package (VASP) \cite{Kresse1,Kresse2}, using the pseudopotential method and a 
plane-wave basis set. The exchange-correlation effects were treated with
generalized gradient-corrected exchange-correlation functionals (GGA) in
the form of Perdew and Wang \cite{Perdew2,Perdew1}, and we adopted the
Vanderbilt ultrasoft pseudopotentials generated by Kresse and Hafner
\cite{Vanderbilt,Kresse0}. The calculations were conducted with a
plane-wave energy cutoff of 20 Ry and 9 special $\vec{k}$ points in
the irreducible part of the two-dimensional Brillouin zone. The
convergence of the total-energy differences (i.e., the binding-energy
differences) between different configurations was checked with 9 and 16
special $\vec{k}$ points using supercells containing five metal layers. We
found that the total-energy differences obtained from calculations with 9
and 16 special $\vec{k}$ points were within a few meV. Extended
convergence checks for Br/Ag(100)-$c(2\times2)$ surfaces suggested that the
use of a cut-off energy of 20 Ry and supercells with a seven-layer Ag slab
separated by a vacuum gap equivalent to seven Ag layers resulted in 
binding-energy differences with errors of only a few meV \cite{SWANG01}. We
therefore believe that the total-energy differences reported in this paper
were within 10 meV of the values that would be obtained with higher cutoff energies, 
more $\vec{k}$ points, or thicker supercells.

The results
obtained from the calculations are summarized in Table~\ref{tab:dft}. The
four-fold hollow configuration is found to be the most stable one. This
is in agreement with the experimental observations. The total energy of
the bridge configuration is higher by 154 meV than the hollow
configuration. The on-top configuration is the least stable, with a
total energy significantly higher than both the hollow and bridge
configurations.

Motivated by the results of these DFT supercell calculations,
we approximated the Ag(100) surface by a two-dimensional sinusoidal
corrugation potential containing $L\times L$ four-fold hollow adsorption sites,
\begin{equation}
U(x,y)=\frac{\Delta}{2}\left[\cos{\left(\frac{2 \pi x}{a}\right)}+\cos{\left(\frac{2 \pi y}{a}\right)}\right],
\label{eq:corrpot}
\end{equation}
where $x\in (0,a L]$ and $y\in (0,a L]$ are the positions on the surface in the two perpendicular lattice directions,
$a=2.889$~{\AA}~\cite{OCKO97} is the two-dimensional lattice constant for Ag(100),
and $\Delta=E_{\rm bridge} - E_{\rm hollow}=E_{\rm top} - E_{\rm bridge}$ gives
the amplitude of the corrugation potential.
This sinusoidal corrugation potential contains the expected rotational symmetries,
but it also has up/down symmetry,
which the actual corrugation potential does not have.
We expect this additional symmetry to have very little effect on the simulations.

The differences between the actual DFT-calculated energies and the approximation, Eq.~(\ref{eq:corrpot}),
with $\Delta=150$~meV are shown in the last column of Table~\ref{tab:dft}.
The value of $\Delta \approx 150$~meV is suggested by $E_{\rm bridge}-E_{\rm hollow}$,
as shown in the third column of Table~\ref{tab:dft}.
The difference between the two potentials is more pronounced
than expected from our earlier $c(2 \times 2)$ calculations \cite{SWANG01}.
The potential well around the four-fold hollow site as obtained by the DFT calculations is much narrower
than that given by the sinusoidal form,
and this difference could be important.
However, in this work we are primarily interested in demonstrating the
general simulation methods required for an off-lattice simulation,
for which the sinusoidal approximation is perfectly adequate.

\begin{figure}[th]
\centerline{\epsfxsize=0.8\textwidth \epsfbox{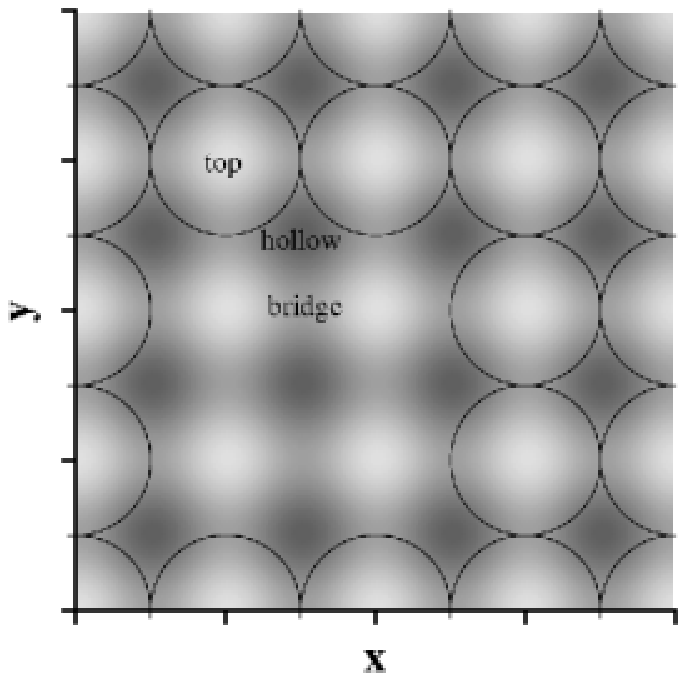}}
\caption{\label{fig:corrulatt}}
\end{figure}

Figure~\ref{fig:corrulatt} shows a grayscale plot of $U(x,y)$.
The dynamic lattice-gas model presented in Refs.~\cite{MITC00A,MITC00C,MITC00B}
implicitly assumed a corrugation potential of this form,
with $\Delta=100$~meV as the nearest-neighbor diffusion barrier.
This somewhat lower amplitude was motivated by the earlier
{\it ab initio} cluster calculations of Ref.~\cite{IGNA98}.
As shown in Table~\ref{tab:dft},
our more recent DFT supercell calculations suggest $\Delta \approx 150$~meV \cite{SWANG01}.

\section{Equilibrium Off-lattice Simulations}
\label{sec:off}

In this section we test the lattice-gas approximation of the Br adlayer
previously assumed in Refs.~\cite{KOPE98,MITC00A,MITC00C,MITC00B}.
Since lattice-gas models assume well-localized surface states
and therefore have only discrete adparticle positions,
they are incapable of reproducing behaviors which depend on lateral
degrees of freedom of the adparticle within the adsorption well.
We therefore present an off-lattice model for the Br adlayer
using the approximate sinusoidal corrugation potential based on the adsorption energies
calculated by DFT in Sec.~\ref{sec:DFT}.
In this off-lattice model,
adparticle positions have continuous degrees of freedom on a two-dimensional surface.

\subsection{Model}

\begin{figure}[th]
\centerline{\epsfxsize=0.8\textwidth \epsfbox{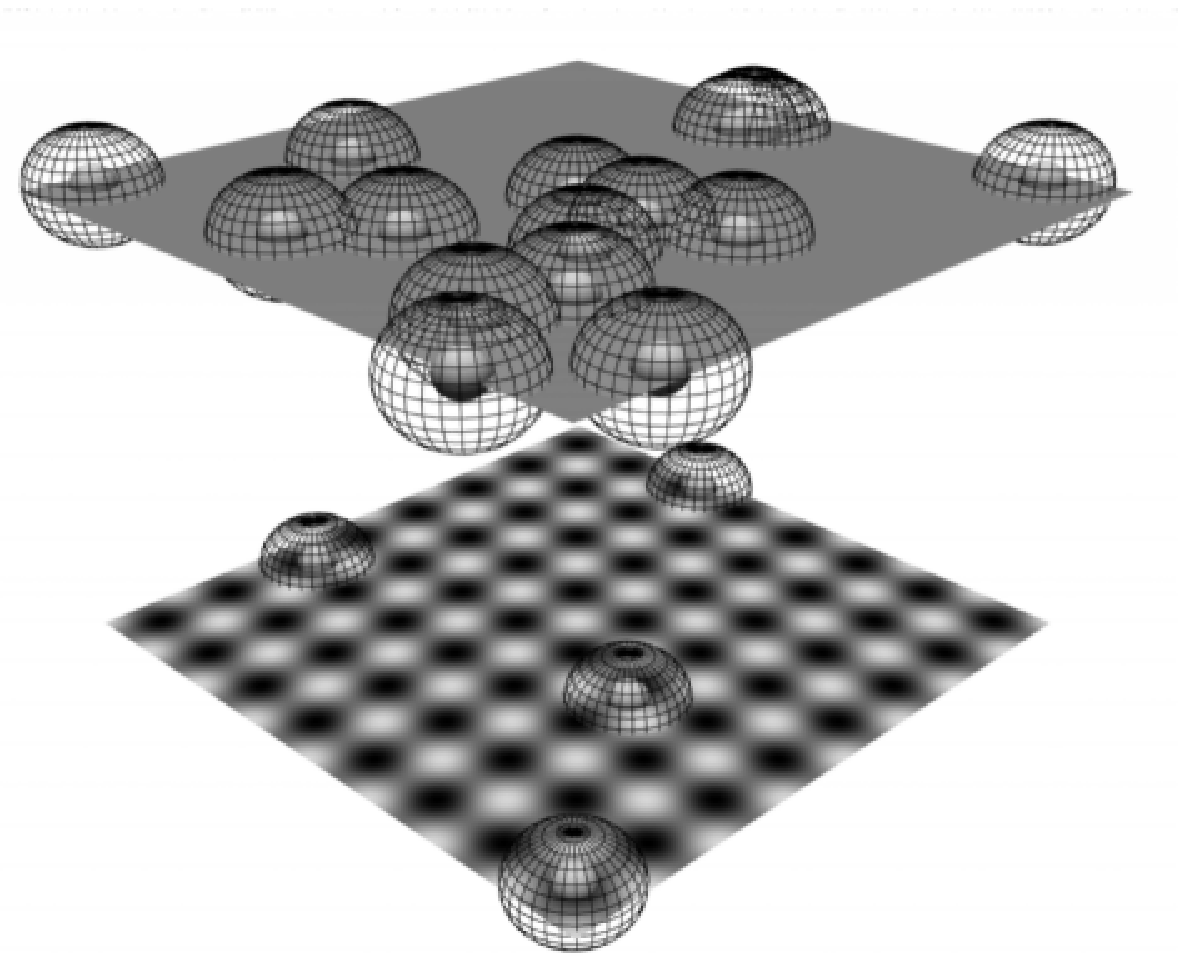}}
\caption{\label{fig:planes}}
\end{figure}

The off-lattice model consists of two parallel planes separated by an unspecified distance,
as shown in Fig.~\ref{fig:planes}.
One plane represents the effective Ag(100) surface (the surface layer),
which includes a corrugation potential for the binding of Br to the surface.
The other plane represents the entire volume of solution near the Ag(100) surface,
from which a Br ion can adsorb within one algorithmic step (the solution layer).
The number of Br ions in the solution layer is $N_{\rm sol}$,
and the number of adsorbed Br is $N_{\rm surf}$.
Both square planes have area $(L a)^2$.
As with the lattice-gas models, periodic boundary conditions are used to eliminate edge effects
in the relatively small systems it is possible to simulate.

Each Br particle has a position $(x,y)$ within one of the layers.
Diffusion moves alter the $(x,y)$ positions of the particles within a layer,
while only adsorption/desorption moves can transfer particles between the layers.
The grand-canonical Hamiltonian for this model is
\begin{equation}
{\mathcal H}=-\frac{1}{2} \sum_{i \ne j } \phi_{i j} + \sum_{i=1}^{N_{\rm surf}} U(x_i,y_i) - \bar{\mu} N_{\rm surf} \; ,
\label{eq:hamil}
\end{equation}
where $i$ and $j$ index adsorbed Br,
$\frac{1}{2}\sum_{i \ne j }$ is a sum over all Br pairs where the factor of $1/2$ is included to prevent double-counting,
$\phi_{i j}$ is the lateral Br-Br interaction energy for the pair $i,j$ within the adlayer,
$\sum_{i=1}^{N_{\rm surf}}$ is a sum over all adsorbed Br particles,
$U(x_i,y_i)$ is the value of the corrugation potential for the $i$-th Br,
and $\bar{\mu}$ is the electrochemical potential.
All energies are measured in units of meV/particle or meV/pair,
and all positions and distances are measured in units of {\AA}.
To simplify the notation, all energy units are simply denoted as meV.
The sign convention used here is such that negative $\phi$ implies
repulsion, more negative $U$ represents stronger binding energy,
and $\bar{\mu}>0$ favors adsorption.

\begin{figure}[th]
\centerline{\epsfxsize=0.8\textwidth \epsfbox{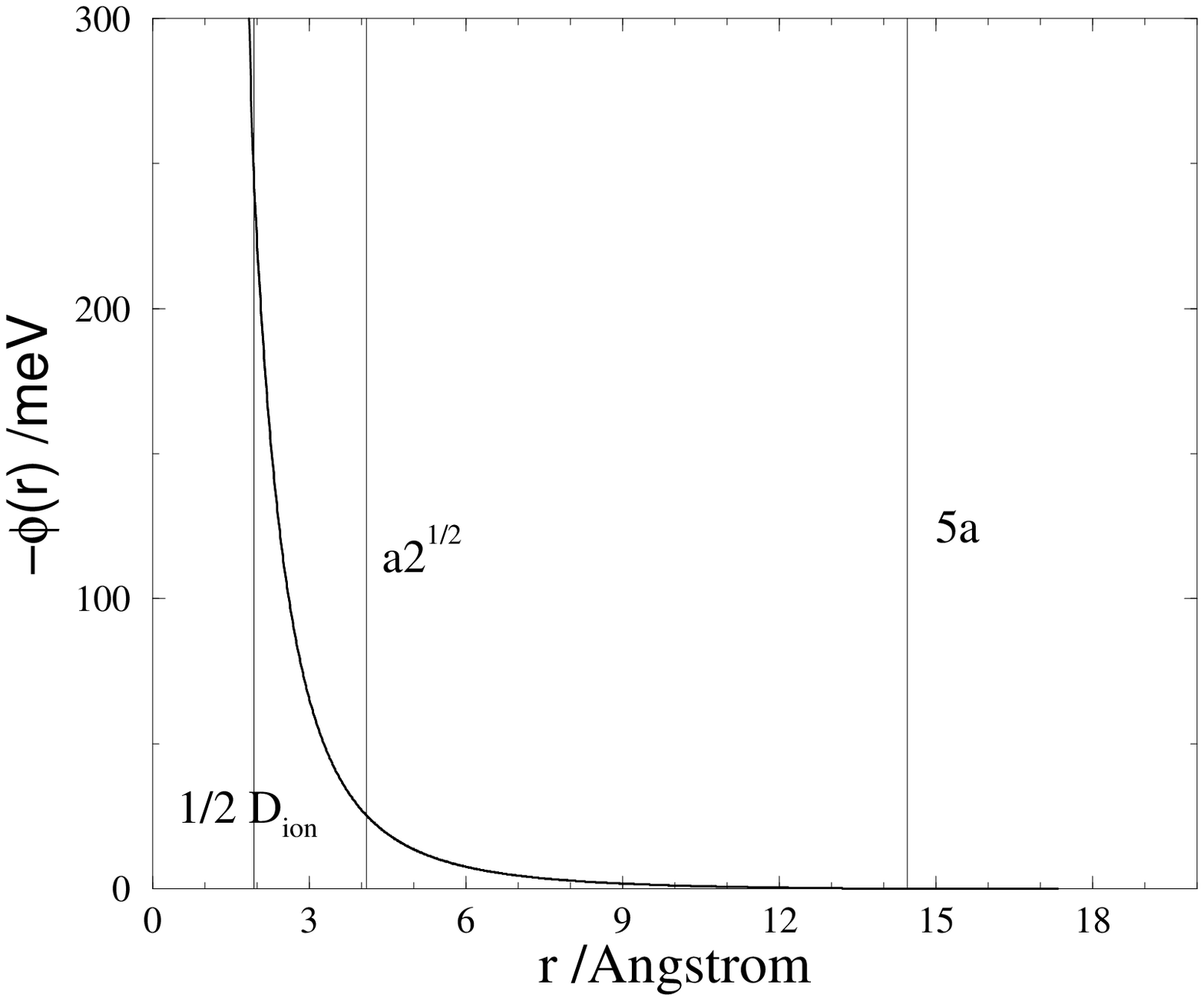}}
\caption{\label{fig:phi}}
\end{figure}

Motivated by the success of the lattice-gas models \cite{KOPE98,MITC00A,MITC00C,MITC00B},
we represent the lateral interactions for large separations as a $1/r^3$ repulsion
and for separations less than the ionic diameter as a truncated Lennard-Jones potential.
The $1/r^3$ potential is most likely associated with both dipole-dipole and lattice-mediated interactions.
Br in the solution layer are non-interacting.
The lateral interaction potential is
\begin{equation}
\phi(r)=
\left\{
\begin{array}{ll}
\phi_{\rm nnn} \left(\frac{a\sqrt{2}}{D_{\rm ion}}\right)^3 - 4 \epsilon \left[\left(\frac{\sigma}{r}\right)^{12}-\left(\frac{\sigma}{r}\right)^{6}\right] - \phi_{\rm nnn} \left(\frac{\sqrt{2}}{5}\right)^3 & r < D_{\rm ion} \\
\phi_{\rm nnn} \left(\frac{a\sqrt{2}}{r}\right)^3 - \phi_{\rm nnn} \left(\frac{\sqrt{2}}{5}\right)^3 & D_{\rm ion} \le r \le 5 a\\
0 & r > 5 a
\end{array}
\right. \; ,
\label{eq:phi}
\end{equation}
where $r$ is the separation between the members of an interacting adsorbed Br pair,
$D_{\rm ion}$ is the ionic diameter of Br ($D_{\rm ion}=2 R_{\rm ion}$ where $R_{\rm ion}=1.94$~{\AA}~\cite{CRSH90}),
$\phi_{\rm nnn}=-26$~meV 
(as determined by fitting lattice-gas simulations to experimental adsorption isotherms in Ref.~\cite{MITC00C})
is the value of $\phi$ at next-nearest neighbor distances ($r=a \sqrt{2}$).
The short-range Lennard-Jones parameters,
$\sigma=R_{\rm ion}$ and $\epsilon=-\phi_{\rm nnn}(a \sqrt{2}/R_{\rm ion})^3/8$,
are found by requiring that $\phi(r)$ be continuous everywhere and that ${\rm d} \phi(r)/{\rm d} r$ be continuous at $r=D_{\rm ion}
$.
Equation~(\ref{eq:phi}) is plotted in Fig.~\ref{fig:phi}.

In the weak-solution approximation, the electrochemical potential $\bar{\mu}$ is related to the temperature,
the concentration of Br ions in solution, and the electrode potential by
\begin{equation}
\bar{\mu}=\bar{\mu}_0+k_B T \ln{\frac{C}{C_0}}-\gamma e E,
\label{eq:mu}
\end{equation}
where $\bar{\mu}_0$ is a reference potential,
$k_B$ is Boltzmann's constant,
$T$ is the absolute temperature ($k_B T=25$~meV at room temperature),
$C$ is the concentration of Br ions in solution,
$C_0$ is a reference concentration of Br ions in solution,
$\gamma$ is the electrosorption valency,
$e$ is the elementary charge unit,
and $E$ is the electrode potential in mV.
Since the off-lattice models considered here involve a form of the solution concentration, $N_{\rm sol}$,
all isotherms require a shift of $\bar{\mu}$ for comparison with the lattice-gas isotherms,
as further explained in the Appendix.

\subsection{Methods}

The equilibrium properties of a system are independent of the dynamical processes.
Thus, we here use the standard Metropolis Monte Carlo (MC)
simulation \cite{METR53} to describe the equilibrium properties of the adlayer at room temperature.
The essence of the MC method is that,
given the current state $I$ of the system,
a new state $F$ is randomly chosen
by making a small change to the state $I$.
Once the energies $E_I$ and $E_F$ of states $I$ and $F$ are calculated using Eq.~(\ref{eq:hamil}),
the new state $F$ is accepted with probability
\begin{equation}
P(F|I)=\min{[1,\exp{(-(E_F-E_I)/k_B T)}]} \; .
\label{eq:metrop}
\end{equation}

We discuss two different types of off-lattice MC simulations.
The first simulations are performed within a fixed-coverage (canonical) ensemble
(i.e., $N_{\rm surf}$ constant).
From these simulations we measure the lateral root-mean-square (rms) displacement
of the adparticles from the four-fold hollow sites as a function of the coverage and the corrugation amplitude $\Delta$.
The second set of simulations is performed within the grand-canonical ensemble (i.e., fixed $\bar{\mu}$),
and the coverage isotherms are calculated for different values of the corrugation amplitude $\Delta$.
Since the fixed-coverage simulations are simpler, we discuss these here.
The grand-canonical MC method \cite{GEORGIEV92} is discussed in detail in the Appendix.

The fixed-coverage simulations begin by placing $N_{\rm surf}=\Theta L^2$ adparticles randomly on the surface.
To simplify the initial MC relaxations, all particles are placed on the same $c(2 \times 2)$ sublattice.
New trial configurations are generated by selecting one of the $N_{\rm surf}$ particles at random.
Call this particle $i$.
A random displacement vector, ${\rm d}\vec{r}$
(which is uniformly distributed within a circle of radius $R_{\rm ion}/4$),
is added to $\vec{r}_i$,
the energy change is calculated,
and the trial configuration is accepted with probability given by Eq.~(\ref{eq:metrop}).

The grand-canonical MC simulations are significantly more complex than the fixed-coverage simulations
due to the added difficulties of satisfying detailed balance under the adsorption/desorption
and lateral diffusion processes.
These difficulties can be overcome by using a ``ghost-particle'' simulation method like that of Ref.~\cite{GEORGIEV92},
which includes both the surface plane and the solution plane shown in Fig.~\ref{fig:planes}.
A full description of the grand-canonical off-lattice MC method is given in the Appendix.

\subsection{Results}

\begin{figure}[th]
\centerline{\epsfxsize=0.8\textwidth \epsfbox{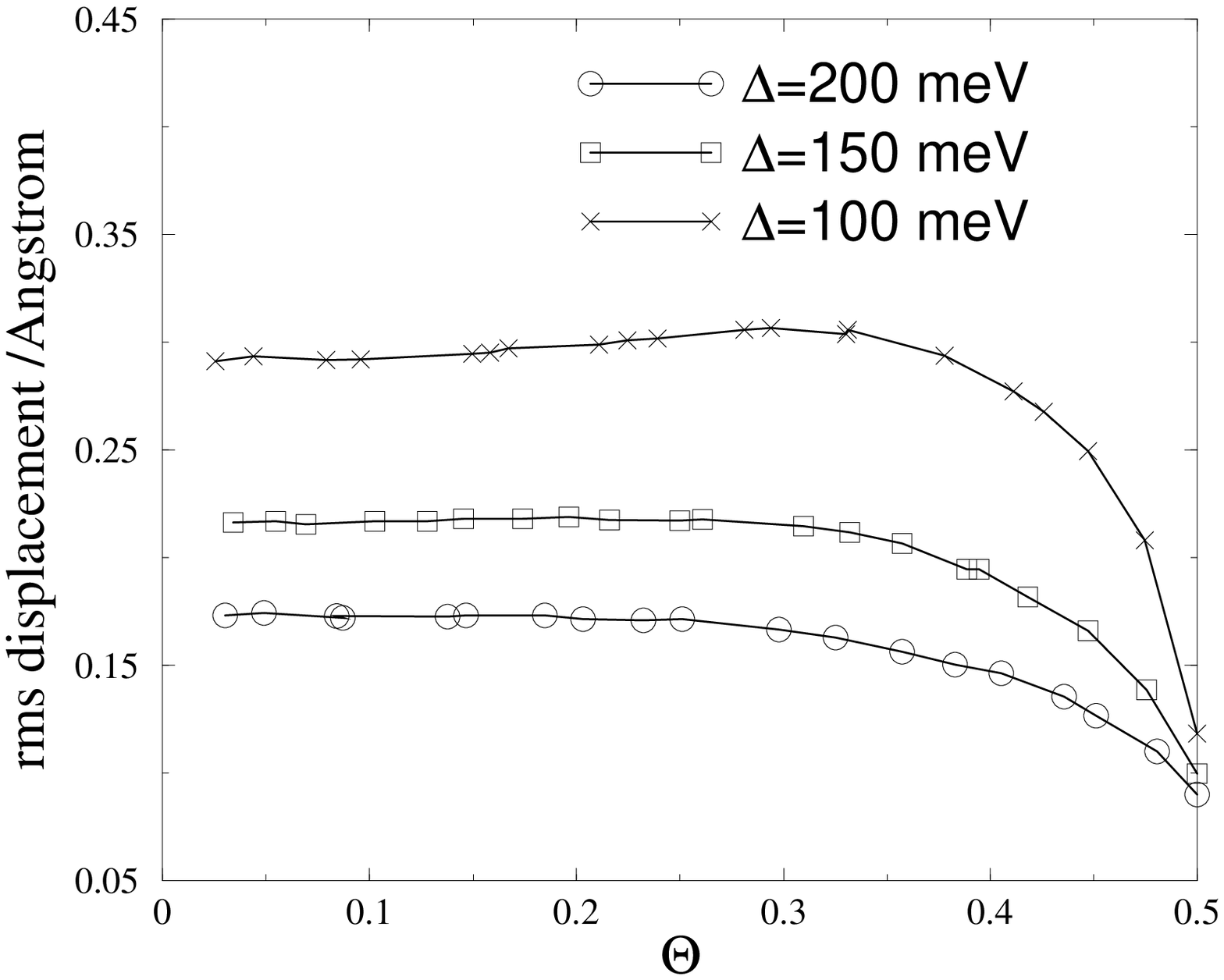}}
\caption{\label{fig:rms}}
\end{figure}

Using canonical MC simulations at room temperature,
the lateral rms displacement of the Br adparticles from the four-fold hollow sites
were measured as a function of coverage and $\Delta$, as shown in Fig.~\ref{fig:rms}.
Two behaviors are obvious.
First, the rms displacement is larger for smaller corrugation amplitudes.
Second, the rms displacement is smallest for coverages above the critical coverage for the $c(2 \times 2)$ structure
($\Theta_{\rm c}\approx 0.37$ \cite{MITC00C}).
This second observation is consistent with results of {\it in situ} X-ray scattering experiments, 
which indicate that the lateral rms displacement is approximately 0.25~{\AA} at low coverages \cite{OCKO01}.
Comparison of these experimental results with Fig.~\ref{fig:rms} suggest that $\Delta \approx 100$~meV.

\begin{figure}[th]
\centerline{\epsfxsize=0.8\textwidth \epsfbox{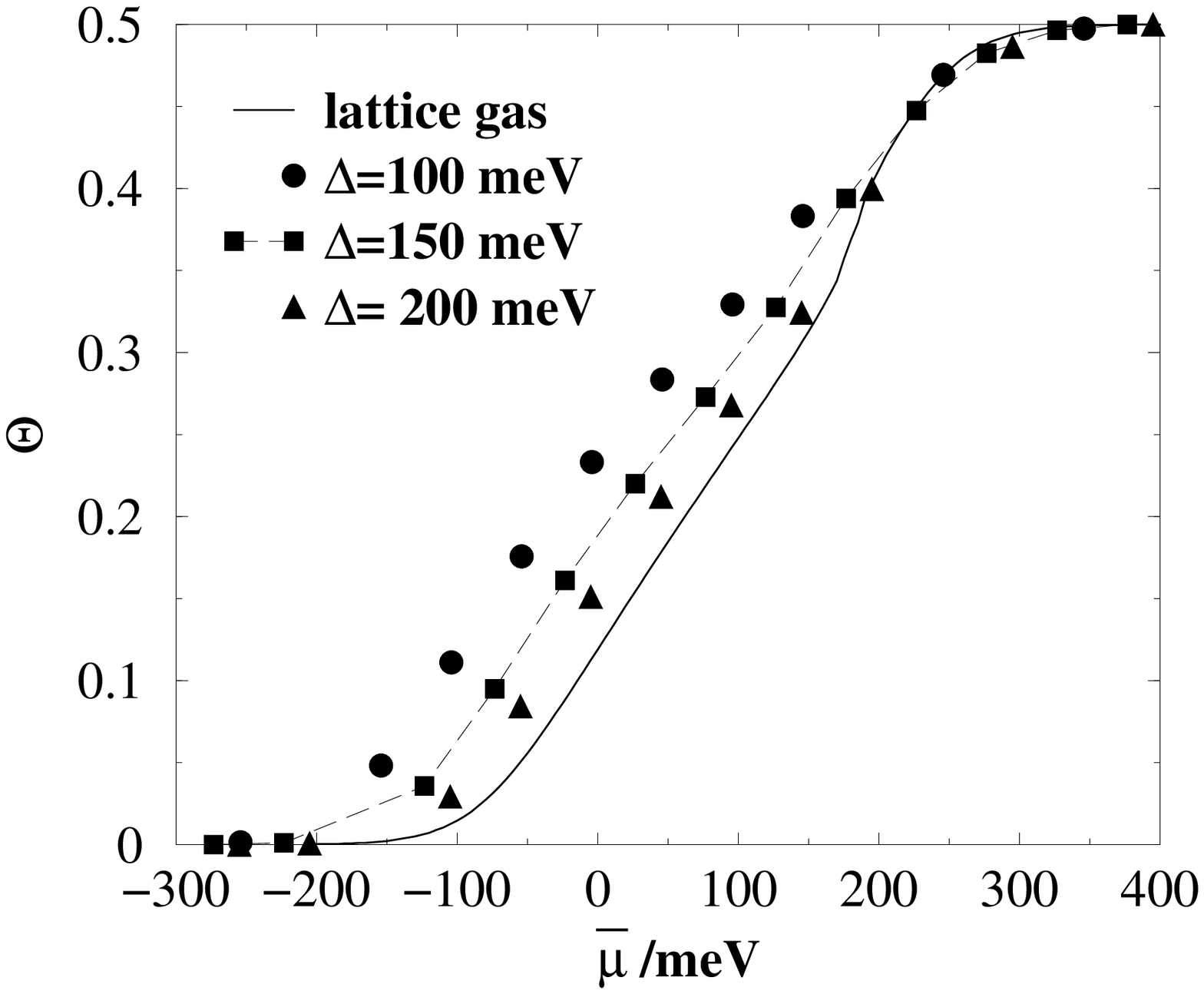}}
\caption{\label{fig:diffbariso}}
\end{figure}

Using the grand-canonical simulation method,
off-lattice equilibrium-coverage isotherms for $\phi_{\rm nnn}=-26$~meV are shown in Fig.~\ref{fig:diffbariso}
for various corrugation amplitudes $\Delta$, along with the corresponding isotherm from the lattice-gas model.
In this figure, the low $\bar{\mu}$ resolution makes identifying the phase transition
between the disordered low-coverage phase and the ordered $c(2 \times 2)$ phases \cite{MITC00A,MITC00C} rather difficult.
The differences are most notable in the disordered phase around $\Theta=1/4$.
For large amplitudes of the corrugation potential ($\Delta \ge 150$~meV),
the disordered region is very linear as seen in both experimental isotherms and the lattice-gas isotherm \cite{MITC00C}.
For lower values of $\Delta$, the isotherm in the disordered region has a pronounced curvature.
No attempt to vary $\phi$ has been made because of the large computational demands such a study would make.
Each of the off-lattice isotherms required approximately 6 weeks of CPU time on a 0.5~GHz PC,
as compared to just a few days of CPU time for the lattice-gas isotherm shown in Fig.~\ref{fig:diffbariso}.
As discussed in Sec.~\ref{sec:DFT}, DFT calculations suggest $\Delta\approx150$~meV \cite{SWANG01};
however, comparison of the rms lateral displacement with experiment suggests $\Delta \approx 100$~meV.
The apparent discrepancy between these two estimates may be resolved by a more accurate approximation
for the shape of the corrugation potential.

Both the lateral rms displacement and the coverage isotherms
indicate that for the reasonable value of $\Delta \approx 150$~meV,
the surface states are well localized near the four-fold hollow sites.
Thus the first approximation made in dynamic Monte Carlo,
well-localized states, is valid for surface diffusion.

\section{Arrhenius Behavior}
\label{sec:arrhenius}

In addition to assuming a lattice-gas treatment of the adlayer,
the previous dynamic Monte Carlo studies of Br/Ag(100) have assumed
that transitions between lattice-gas states are well
described by thermally activated stochastic barrier hopping \cite{MITC00A,MITC00C,MITC00B}.
In particular, the Arrhenius form for the transition rates was assumed:
\begin{equation}
R( F | I )= \nu_{\rm eff} \exp{\left( - \Delta_{\rm eff}/k_B T\right)} \; ,
\label{eq:arhenrate}
\end{equation}
where $R( F | I )$ is the probability per unit time of undergoing
a transition between a lattice-gas state $I$ and a lattice-gas state $F$,
$\nu_{\rm eff}$ is an effective attempt frequency,
and $\Delta_{\rm eff}$ is the effective free-energy barrier.
In general, $\Delta_{\rm eff}$ has a dependence on the energies of the states $I$ and $F$,
a dependence on the bare barrier of the process,
which generally is different for diffusion and adsorption/desorption,
and it also includes a dependence on entropic effects,
such as the accessibility of the intermediate transition state.

To investigate the validity of the Arrhenius assumption,
we must perform dynamic simulations which consider
both the continuous particle positions and the kinetic energy.
Conventional dynamic Monte Carlo ignores the kinetic energy.
The ideal approach would be a full molecular-dynamics treatment
of the substrate atoms, adsorbed Br ions,
and all of the atoms in the solution.
However, such a simulation would require not only knowledge of the very complex interaction potentials,
but also an enormous computational effort.
Since the system is dominated by thermally activated barrier hopping,
it is unlikely that conventional molecular-dynamics simulations could reach the necessary long time scales.
The full simulation might be feasible within the Hyperdynamics Method of Voter \cite{VOTE97},
although Hyperdynamics assumes Arrhenius type behavior.
Thus, we here present a simplified model of the dynamics which still includes
most of the essential behavior, such as kinetic energy (ignored in MC dynamics) and thermal noise.
However, as with the lattice-gas models,
there is no explicit treatment of the solution and substrate atoms.

We investigate the validity of the Arrhenius assumption by performing
Langevin simulations of a single Br particle on the surface.
The behavior of only a single particle is used to avoid lateral Br-Br interactions,
thus making the calculations completely general for any system with a corrugation potential
well approximated by Eq.~(\ref{eq:corrpot}).
We then analyze the single-particle trajectory in different parameter regimes
to quantify the Arrhenius behavior and deviations from it.

\subsection{Langevin Simulations}

To simplify the simulation, we assume a static substrate,
where the average positions of the Ag atoms give rise to the corrugation potential of Eq.~(\ref{eq:corrpot}),
and the fluctuations of the Ag atoms about their average positions are included
by introducing drag and random-noise terms which represent collisions with phonons in the Ag crystal.
The presence of water is replaced by additional drag and random-noise terms,
which are caused by collisions with water molecules.

We only investigate the lateral diffusion of a single adparticle on the surface,
and we therefore ignore adsorption/desorption processes.
The drag and noise terms from both substrate and solution are combined
into a Langevin equation of motion \cite{ALLEN87} governing the lateral motion of the Br particle within the surface layer
\begin{equation}
m \dot{\vec{v}}(t)= - \vec{\bigtriangledown} U(\vec{x}(t)) -\alpha \vec{v}(t) + \vec{\eta}(t) \; ,
\label{eq:langevin}
\end{equation}
where $m$ is the effective mass of a Br ion,
$t$ is time,
$\dot{\vec{v}}(t)$ is the time derivative of the velocity $\vec{v}(t)$ of the Br adparticle at time $t$,
$\vec{x}(t)$ is the two-dimensional position of the adparticle at time $t$,
$- \vec{\bigtriangledown} U(\vec{x}(t))$ is the force on the adparticle due to the corrugation potential,
$\alpha$ is the drag coefficient,
and $\vec{\eta}_i(t)$ is a Gaussian white random force.
For simplicity, we assume that $m$ is equal to the mass of a Br atom,
and for convenience the values of $\alpha$ are reported in units of
$80 \times 10^{-15}$~kg/s,
corresponding to the system of dimensionless units used in the numerical simulations.

The Gaussian random force has the following properties \cite{ALLEN87}:
\begin{equation}
\begin{array}{c}
\langle \eta_x(t) \rangle = \langle \eta_y(t) \rangle =0 \\
\langle \eta_x(t) \eta_y(t') \rangle =0 \\
\langle \eta_x(t) \eta_x(t')\rangle =\sigma_{\rm G}^2 \delta(t-t') \\
\langle \eta_y(t) \eta_y(t')\rangle =\sigma_{\rm G}^2 \delta(t-t') \; ,
\end{array}
\label{eq:noiseprops}
\end{equation}
where the subscripts denote the $x$ and $y$ components of the random force,
$t$ and $t'$ are two arbitrary times,
$\langle \cdot \rangle$ represents an ensemble or time average,
$\sigma_{\rm G}^2$ is the variance of a Gaussian distribution,
and $\delta(\cdot)$ is the Dirac delta function.
The width of the Gaussian noise distribution is related to the friction coefficient $\alpha$
and the temperature $T$ by the fluctuation-dissipation theorem \cite{ALLEN87},
which requires that
\begin{equation}
\sigma_{\rm G}=\sqrt{2 k_B T \alpha} \; .
\label{eq:flucdisp}
\end{equation}

Equation~(\ref{eq:langevin}) is solved iteratively by a first-order integration method,
\begin{equation}
\vec{v}(t+\Delta t)=\vec{v}(t) - \vec{\bigtriangledown} U(\vec{x}(t)) \Delta t/m -\alpha \vec{v}(t)\Delta t/m + \vec{\eta}(t)\sqrt
{\Delta t}/m \; ,
\label{eq:velmotion}
\end{equation}
where $\Delta t$ is the magnitude of the discrete time step,
here taken as $\Delta t=1.66\times 10^{-15}$~s.
The simulations were not numerically stable for $\Delta t$ significantly larger than this value.
For smaller $\Delta t$, no significant differences were seen.
The $\sqrt{\Delta t}$ factor in the term $\vec{\eta}_i(t)\sqrt{\Delta t}/m$
is a commonly known result of stochastic noise integration \cite{ALLEN87}
and essentially corresponds to the noise performing a random walk as a function of time.
The new positions are given by
\begin{equation}
\vec{x}(t+\Delta t)=\vec{x}(t)+\vec{v}(t)\Delta t \; .
\label{eq:posmotion}
\end{equation}

\subsection{Trajectory Results}

\begin{figure}[th]
\centerline{\epsfxsize=0.8\textwidth \epsfbox{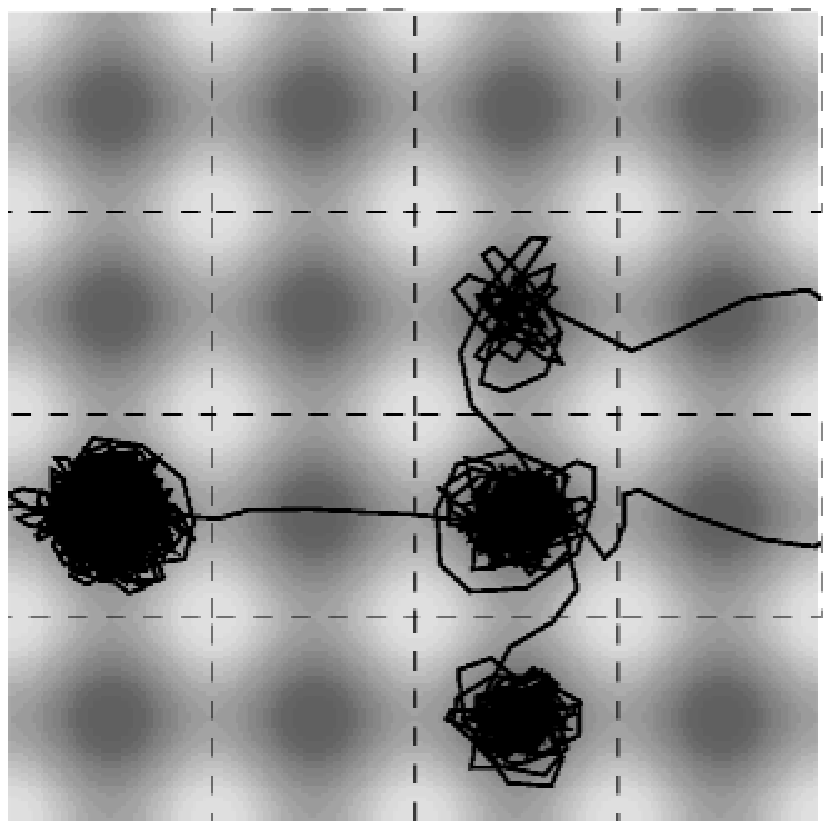}}
\caption{\label{fig:traj}}
\end{figure}

Langevin simulations of a single-particle trajectory were performed at room temperature.
We define the crossing time, $\tau$, as the time spent within the current unit cell of the Ag(100) lattice
before moving into another adsorption well.
The particle was placed on the surface, and after throwing away the first few crossings,
a time sequence of $\tau$ was measured including 10,000 crossings.
A typical trajectory is shown in Fig.~\ref{fig:traj},
superimposed on the corrugation potential and a grid indicating the unit cells.
The bridge site is seen to be the dominant location for hopping.
Though it is difficult to see in the static image of Fig.~\ref{fig:traj},
three behaviors are easily seen in dynamic images of the trajectories:
1) Arrhenius-like single, independent hops between cells,
2) crossing and immediate recrossing back into the previous cell,
and 3) crossing multiple bridge sites in a very short time, which we call ``running.''
The dynamic MC methods used in the lattice-gas simulations included only Arrhenius-like crossings.

\begin{figure}[th]
\centerline{\epsfxsize=0.8\textwidth \epsfbox{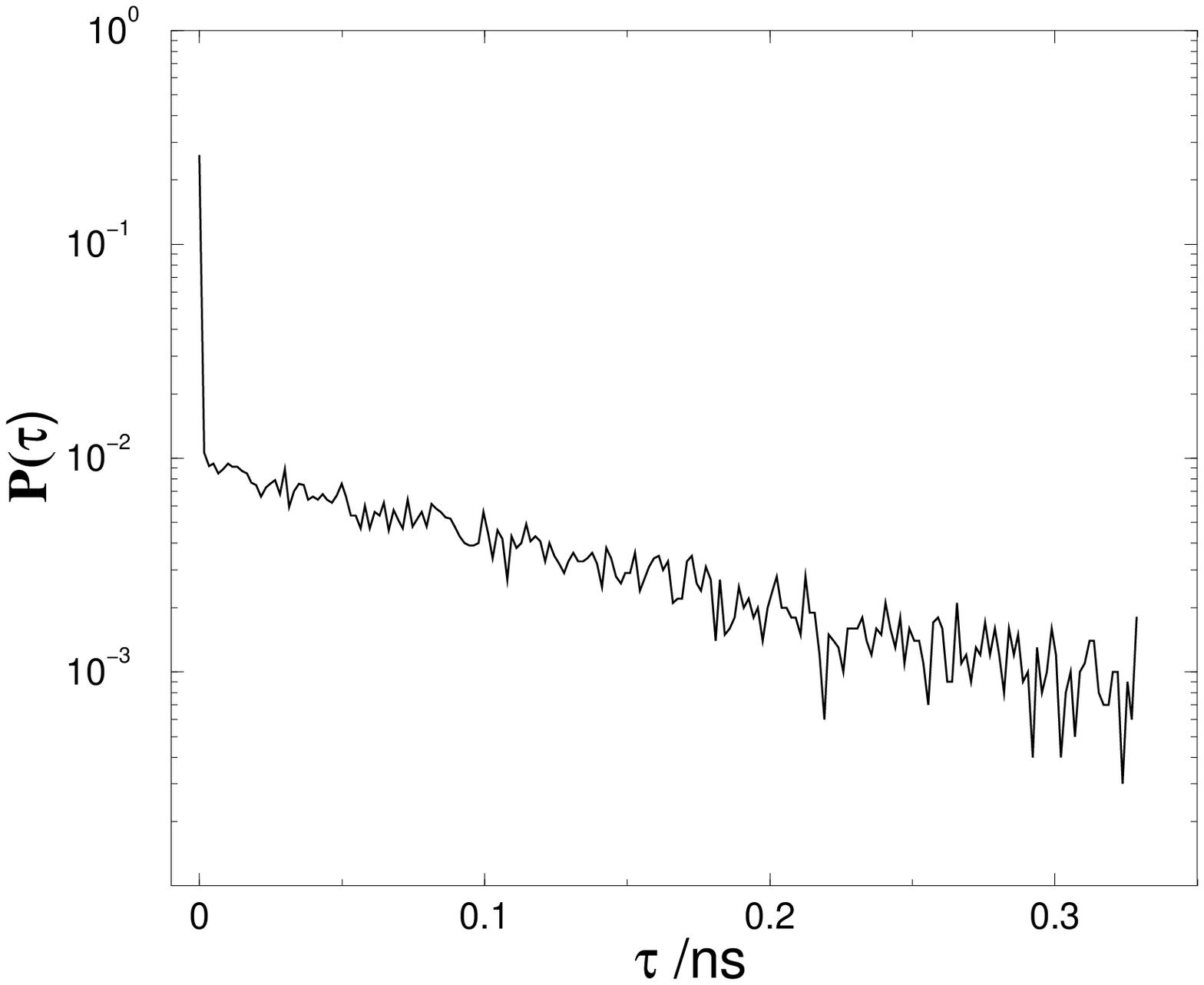}}
\caption{\label{fig:taudist}}
\end{figure}

The Arrhenius behavior and the small-$\tau$ deviations from it can be seen in the probability density of $\tau$,
shown in Fig.~\ref{fig:taudist}.
The Arrhenius behavior is observed as an exponential tail in the distribution,
which occurs for sufficiently large $\tau$,
regardless of the value of the damping $\alpha$ ($\alpha > 0$) or the corrugation amplitude $\Delta$ ($\Delta > k_B T$).
However, a large number of crossings occur in the short-$\tau$ regime.
Furthermore, in the weakly damped limit (very small $\alpha$),
the short-$\tau$ crossings should be dominated by running,
while in the strongly damped limit (very large $\alpha$),
crossing/recrossing behavior should dominate the small-$\tau$ crossings.

\begin{figure}[th]
\centerline{\epsfxsize=0.95\textwidth \epsfbox{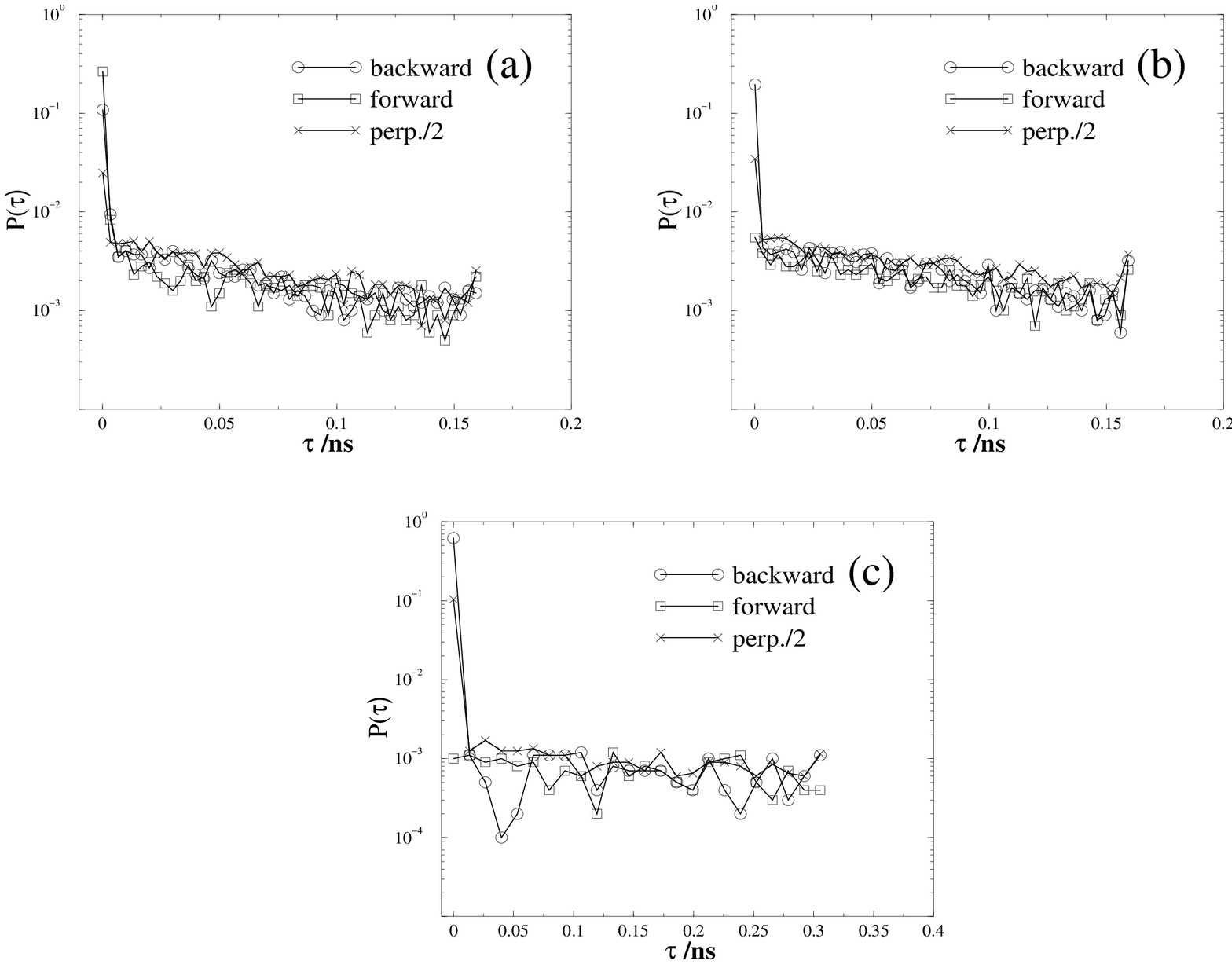}}
\caption{\label{fig:dirprob}}
\end{figure}

Before examining the large-$\tau$ Arrhenius behavior in greater detail,
we quantify the amount of the three crossing behaviors
by examining the probability to cross in a particular direction,
relative to the direction of the previous crossing.
When the damping is small (the thermalization time is long),
then a particle which has sufficient kinetic energy to hop one barrier
will also have sufficient kinetic energy to immediately hop a second barrier.
This is the cause of running.
In contrast, if the thermalization time is short (large damping),
then the small-$\tau$ crossings should correspond to recrossing back into the previous cell.
Figure.~\ref{fig:dirprob} shows the directional crossing probabilities
for $\Delta=150$~meV and various values of $\alpha$ as functions of $\tau$.
For $\alpha=1$, the small-$\tau$ regime is dominated by running,
but for $\alpha=10$ and $\alpha=100$, the small-$\tau$ regime is dominated by crossing/recrossing events.

\begin{figure}[th]
\centerline{\epsfxsize=0.95\textwidth \epsfbox{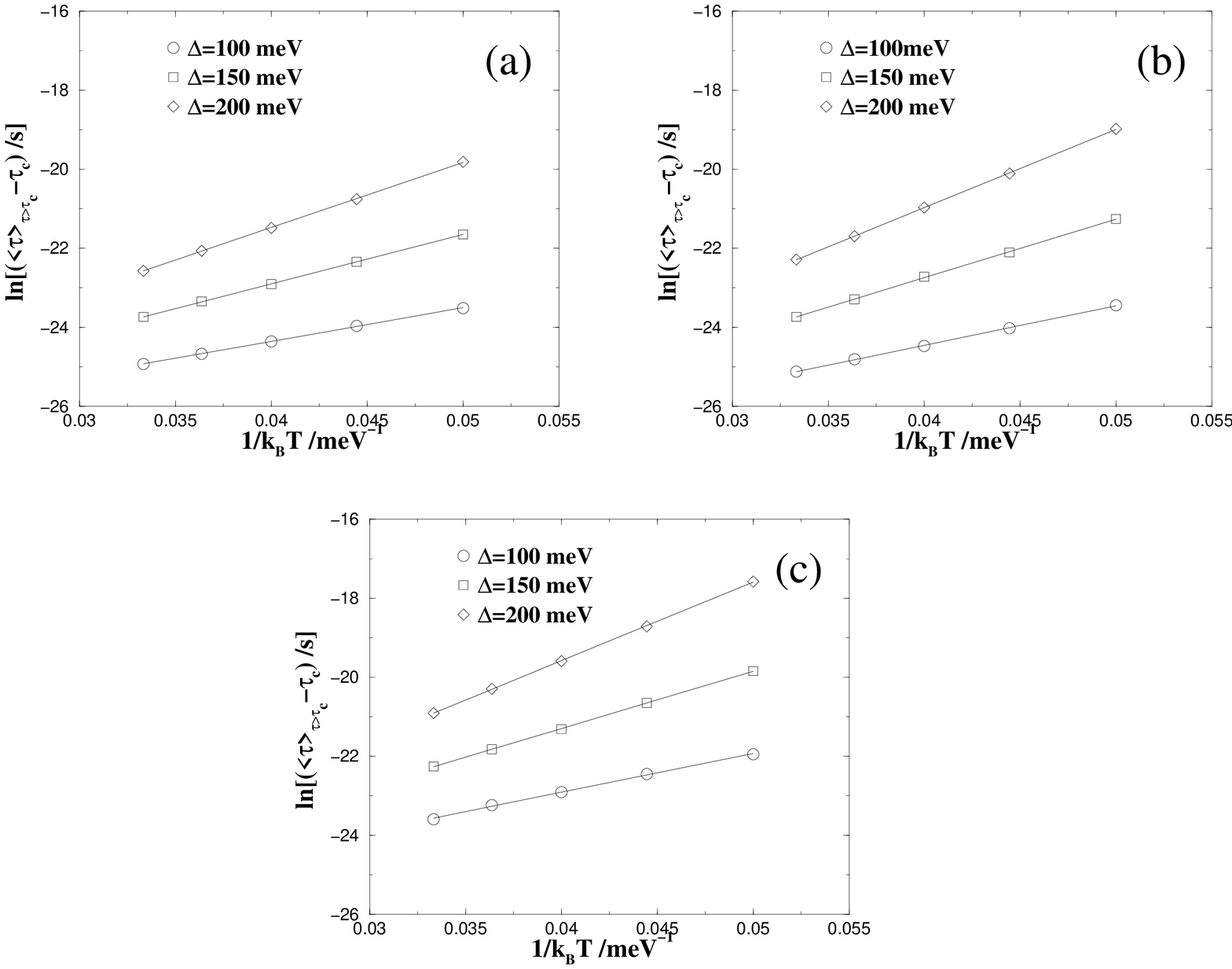}}
\caption{\label{fig:nubarr}}
\end{figure}

To further elucidate the long-time Arrhenius behavior,
we next examine the temperature dependence of the crossing rate.
Because the exponential behavior occurs only for large crossing times,
we only examine crossing times larger than a minimum cutoff, $\tau_c=8.3$~ps.
We assume the standard Arrhenius form for the crossing rate given in Eq.~(\ref{eq:arhenrate}).
Arrhenius behavior assumes a Poisson process with a constant rate
which has the probability distribution $P(\tau) = R \exp{[-R (\tau-\tau_c)]}$, for $\tau \ge \tau_c$.
After integrating $\tau P(\tau)$ for $\tau \ge \tau_c$,
we obtain the expression
\begin{equation}
\langle \tau \rangle_{\tau \ge \tau_c}=\tau_c+R^{-1}=\tau_c+\nu_{\rm eff}^{-1} \exp{[\Delta_{\rm eff}/(k_B T)]} \; ,
\label{eq:avgtau}
\end{equation}
where $\langle \tau \rangle_{\tau \ge \tau_c}$ represents the average crossing time such that $\tau \ge \tau_c$.
From this expression, both the effective barrier and the attempt frequency can be found
from a plot of $\ln{\left[ \langle \tau \rangle_{\tau \ge \tau_c} -\tau_c)\right]}$ vs $(k_B T)^{-1}$,
as shown in Fig~\ref{fig:nubarr}.
In this plot, the intercept is $-\ln{(\nu})$, and the slope is $E_{\rm barr}$.
A summary of the effective Arrhenius parameters is shown in Table~\ref{tab:effbarr} and Table~\ref{tab:effnu}.
Significant deviations from the assumption made in Refs.~\cite{MITC00C,MITC00A,MITC00B}
that $E_{\rm barr}=\Delta$ are seen for the weaker damping.
Without knowledge of the actual value of $\alpha$, no prediction of $\nu_{\rm eff}$ can be made;
however, we note that for all values of $\alpha$ and $\Delta$ studied here,
$\nu_{\rm eff}$ is on the order of 10$^{12}$~s$^{-1}$.

As a final comment, the repulsive lateral interactions between Br in the adlayer
should alter the crossing behavior in the presence of other Br particles.
However, these effects have yet to be studied.

\section{Conclusions}
\label{sec:conc}

To investigate the validity of the lattice-gas approximation
or well-localized surface-state assumption,
we have constructed an off-lattice model using a corrugation potential
approximated on the basis of Density Functional Theory (DFT) surface-binding-energy calculations.
Equilibrium Monte Carlo (MC) simulations in both the canonical (fixed coverage)
and grand-canonical (fixed $\bar{\mu}$) ensembles were performed.
Both the rms lateral displacement from the binding sites
and the coverage isotherms suggest that for the most reasonable value
of the corrugation potential, $\Delta \approx 150$~meV,
the surface states are well localized,
thus justifying the previous lattice-gas models.

The stochastic Arrhenius transition rate assumption was also tested using
Langevin simulations of single-particle trajectories.
These trajectories are general for any system where the corrugation potential
is well approximated by the sinusoidal form used in this paper.
For long time-scale transitions,
the transition rate was found to be well described by the Arrhenius rate.
However, crossing/recrossing and running behaviors were observed for short-time transitions,
suggesting that future dynamic Monte Carlo simulations should include such behaviors.
Unfortunately, the value of the phenomenological damping parameter is unknown for Br/Ag(100),
and it is therefore unknown whether the short-time transitions for the physical system
fall into the crossing/recrossing or running regimes.

Future theoretical, computational, and experimental efforts should concentrate
on determining the damping parameter as well as further justification of the
dynamic Monte Carlo models.
Finally, we point out that other processes, such as adsorption/desorption and molecular reorientation,
can be studied with similar off-lattice models,
where the two-dimensional corrugation potential is generalized to three dimensions.

\section*{Acknowledgments}

The authors would like to thank G.\ Brown, Th.\ Wandlowski, and B.~M.\ Ocko for useful discussions
and G.\ Brown for comments on the manuscript.
This research was funded by the U.S. National Science Foundation through grant No. DMR-9981815,
and by Florida State University through the Center for Materials Research and Technology
and the School of Computational Science and Information Technology.

\appendix
\section*{Appendix}

In this Appendix we discuss the Monte Carlo methods
used to simulate the equilibrium properties of the off-lattice model.
This method is identical to the ghost-particle method of Ref.~\cite{GEORGIEV92}.
However, that reference contains several errors,
and the correct details of the method are given here.

The grand-canonical partition function is
\begin{equation}
\mathcal{Z} = \sum_{N=0}^{\infty} \exp{\left( \beta \mu N  \right)} \mathcal{Z}_N \; ,
\label{eq:gcpart}
\end{equation}
where $N$ is the number of particles in the adlayer,
$\mathcal{Z}_0=1$, and
\begin{equation}
\mathcal{Z}_N = h^{-Nd} (N ! )^{-1}
\int {\rm d}^d\vec{p}_1 \cdots {\rm d}^d\vec{p}_N {\rm d}^d\vec{r}_1 \cdots {\rm d}^d\vec{r}_N
\exp{\left( -\beta \mathcal{H}_N \right)}\; ,
\label{eq:configpart}
\end{equation}
where $\vec{p}_i$ and $\vec{r}_i$ are the two-dimensional momentum and position, respectively,
of the $i$th particle,
$\mathcal{H}_N$ is the Hamiltonian of the $N$-particle system,
including both kinetic and potential energies,
$d$ is the spatial dimension,
$\beta=1/k_B T$,
and $h$ is Planck's constant.
For this work, $d=2$,
and we make no distinction between $\mu$ and $\bar{\mu}$.

As $N \rightarrow \infty$, the typical distance between neighboring particles goes to zero.
In this limit, the interaction energy will go to infinity (see Chapter~4),
and thus the exponential factor in Eq.~(\ref{eq:configpart}) will go to zero.
Thus, we can consider some maximum number of particles, $M$,
such that the probability of having a configuration with $N>M$ is negligible.
This leads to the approximation of Eq.~(\ref{eq:gcpart}),
\begin{equation}
\mathcal{Z} \approx \sum_{N=0}^{M} \exp{\left( \beta \mu N  \right)} \mathcal{Z}_N \; .
\label{eq:approxgcpart}
\end{equation}
However, this is the partition function for a canonical ensemble
describing the system consisting of the surface layer and the solution together.
The remainder of this Appendix will be concerned with constructing such
a canonical ensemble which correctly approximates a grand-canonical ensemble for the surface layer.

Consider a system of $M$ particles with continuous momentum and position degrees of freedom.
We add the additional discrete degree of freedom, $c$, which is 1 for particles in the surface layer
and 0 for particles in the solution layer.
Particles in the solution layer are sometimes referred to as ``ghost'' particles.
This system has $2dM$ continuous degrees of freedom
and $M$ discrete degrees of freedom.
The number of particles in the adlayer is then
\begin{equation}
N = \sum_{i=1}^M c_i \; ,
\end{equation}
where $c_i$ denotes whether the $i$th particle is in the solution (0) or the surface (1) layer.
The Hamiltonian for this system is
\begin{equation}
\mathcal{H}_{\rm tot} =
\sum_{i=1}^M \frac{\vec{p}_i \; ^2}{2m} +
\sum_{i=1}^{M-1} \sum_{j=i+1}^{M} U_{\rm int}(|\vec{r}_i-\vec{r}_j|) c_i c_j +
\sum_{i=1}^M U_{\rm ext}(\vec{r}_i) c_i +
\Phi(N,M,\mu,\beta,V) \; ,
\label{eq:htot}
\end{equation}
where $m$ is the effective mass of a particle,
$U_{\rm int}(|\vec{r}_i-\vec{r}_j|)$ is the interaction potential between two particles,
$U_{\rm ext}(\vec{r}_i)$ is an external potential, here equivalent to the corrugation potential,
and $\Phi(N,M,\mu,\beta,V)$ is a correction factor which we will now derive.
Note also that particles in the solution layer have only kinetic energy.

The partition function for this system is
\begin{equation}
\mathcal{Z}_{\rm tot} = h^{-Md} (M !)^{-1}
\sum_{c_1=0}^1 \sum_{c_2=0}^1 \cdots \sum_{c_M=0}^1
\int {\rm d}^d\vec{p}_1 \cdots {\rm d}^d\vec{p}_M {\rm d}^d\vec{r}_1 \cdots {\rm d}^d\vec{r}_M
\exp{\left( -\beta \mathcal{H}_{\rm tot} \right)} \; .
\label{eq:totpart}
\end{equation}
The correction factor in Eq.~(\ref{eq:htot}) can be found by equating Eq.~(\ref{eq:totpart}) and Eq.~(\ref{eq:approxgcpart}).
We first rewrite Eq.~(\ref{eq:totpart})
\begin{equation}
\begin{array}{ll}
\mathcal{Z}_{\rm tot} = & h^{-Md} (M !)^{-1} \sum_{N=0}^M \left( \begin{array}{c} M \\ N \end{array} \right)
\int {\rm d}^d\vec{p}_1 \cdots {\rm d}^d\vec{p}_M {\rm d}^d\vec{r}_1 \cdots {\rm d}^d\vec{r}_M \\
& \times {\rm exp}\left( -\beta \left[ \sum_{i=1}^M \frac{\vec{p}_i \; ^2}{2m} +
\sum_{i=1}^{N-1} \sum_{j=i+1}^{N} U_{\rm int}(|\vec{r}_i-\vec{r}_j|) c_i c_j \right. \right. \\
& \left. \left. + \sum_{i=1}^N U_{\rm ext}(\vec{r}_i) c_i + \Phi(N,M,\mu,\beta,V) \right] \right) \; ,
\end{array}
\label{eq:totpart2}
\end{equation}
which  can be rewritten in this way for the following reasons:
\begin{enumerate}
\item The particles can always be renumbered such that the first $N$ are surface-layer particles.
This means it is superfluous to write $c_i$,$c_j$ in the surface sums,
since they are unity by definition.
\item The number of ways to assign the occupation variables $c_i$ for $N$ surface particles is $\left( \begin{array}{c} M \\ N \end
{array} \right)$.
\item The number of terms in Eq.~(\ref{eq:totpart}) and Eq.~(\ref{eq:totpart2}) are both $2^M$.
\end{enumerate}

After factoring out the integrals over the momenta and positions of the $M-N$ solution layer particles,
Eq.~(\ref{eq:totpart2}) becomes
\begin{equation}
\begin{array}{ll}
\mathcal{Z}_{\rm tot} = & \sum_{N=0}^M h^{-Md} (M !)^{-1} \left( \int {\rm d}^d \vec{r} \right)^{M-N}
\left( \int {\rm d}^d \vec{p} \exp{\left( -\beta \vec{p} \; ^2 / 2 m \right)} \right)^{M-N}
\left( \begin{array}{c} M \\ N \end{array} \right) \\
& \times \int {\rm d}^d\vec{p}_1 \cdots {\rm d}^d\vec{p}_N {\rm d}^d\vec{r}_1 \cdots {\rm d}^d\vec{r}_N
{\rm exp}\left( -\beta \left[ \mathcal{H}_N + \Phi(N,M,\mu,\beta,V) \right] \right) \; .
\end{array}
\label{eq:totpart3}
\end{equation}
This equation can be easily rewritten,
\begin{equation}
\begin{array}{lll}
\mathcal{Z}_{\rm tot} & = & \sum_{N=0}^M
\left[ \left( V \right)^{M-N}
\left( \frac{2 \pi m k_B T}{h^2} \right)^{d(M-N)/2}
 \frac{{\rm exp}\left( -\beta \Phi(N,M,\mu,\beta,V) \right)}{(M-N)!} \right] \\
& & \times \left[ h^{-Nd} (N!)^{-1} \int {\rm d}^d\vec{p}_1 \cdots {\rm d}^d\vec{p}_N {\rm d}^d\vec{r}_1 \cdots {\rm d}^d\vec{r}_N
{\rm exp}\left( -\beta \mathcal{H}_N \right) \right]\\
& = & \sum_{N=0}^M
\left[ \left( V \right)^{M-N}
\left( \frac{2 \pi m k_B T}{h^2} \right)^{d(M-N)/2}
 \frac{{\rm exp}\left( -\beta \Phi(N,M,\mu,\beta,V) \right)}{(M-N)!} \right] \mathcal{Z}_N \; .
\end{array}
\label{eq:totpart4}
\end{equation}
Equating Eq.~(\ref{eq:totpart4}) and Eq.~(\ref{eq:approxgcpart}) now yields
\begin{equation}
\left( \frac{V}{\Lambda^d} \right)^{M-N}
\frac{{\rm exp}\left( -\beta \Phi(N,M,\mu,\beta,V) \right)}{(M-N)!}
= {\rm exp}\left( \beta \mu N \right) \; ,
\end{equation}
where
\begin{equation}
\Lambda = \left( \frac{h^2}{2 \pi m k_B T} \right)^{1/2} \; ,
\end{equation}
is the thermal wavelength.

The value of the correction factor is thus
\begin{equation}
\Phi(N,M,\mu,\beta,V) = -\mu N + k_B T \left[ (M-N) {\rm ln}{\left( \frac{V}{\Lambda^d} \right)}  -{\rm ln}{\left( (M-N)!  \right)}
 \right] \; .
\end{equation}
This correction term may be included directly into the Hamiltonian,
or we may redefine the zeros of the energy and chemical potential,
\begin{equation}
\mathcal{H}_{\rm tot}'=\mathcal{H}_{\rm tot} - k_B T M {\rm ln}{\left( \frac{V}{\Lambda^d} \right)} \; ,
\end{equation}
and
\begin{equation}
\mu' = \mu + k_B T {\rm ln}{\left( \frac{V}{\Lambda^d} \right)} \; .
\end{equation}
This yields the effective Hamiltonian
\begin{equation}
\mathcal{H}_{\rm tot}' =
\sum_{i=1}^M \frac{\vec{p}_i \; ^2}{2m} +
\sum_{i=1}^{M-1} \sum_{j=i+1}^{M} U_{\rm int}(|\vec{r}_i-\vec{r}_j|) c_i c_j +
\sum_{i=1}^M U_{\rm ext}(\vec{r}_i) c_i
-\mu'N -k_B T {\rm ln}{\left( (M-N)!  \right)} \; .
\label{eq:htotprime}
\end{equation}
Monte Carlo energy differences are then calculated with this effective Hamiltonian.
There is no need to treat the particle momenta
in a system with no momentum-dependent forces,
and we do not change the kinetic energy during the Monte Carlo procedure.

The Monte Carlo proceeds in the following way:
\begin{enumerate}
\item Choose one of the $M$ particles at random.  Call the chosen particle $i$.
\item Randomly choose to diffuse or adsorb/desorb particle $i$.
The relative probability of choosing diffusion or adsorption/desorption is not particularly important,
and for simplicity, the probability of choosing diffusion is set to $1/2$.
        \begin{itemize}
        \item If diffusion is chosen, choose the new $\vec{r}_{i} \; '=\vec{r}_i+{\rm d}\vec{r}$,
                where $\vec{r}_i \; '$ is the new position
                and ${\rm d}\vec{r}$ is a uniformly chosen displacement within a $d$-dimensional
                sphere of radius $R$.
        \item If adsorption/desorption is chosen, change the layer occupation variable, $c_i$,
                such that if $c_i=0$, then $c_i'=1$, and if $c_i=1$, then $c_i'=0$.
        \end{itemize}
\item Calculate the energy change using Eq.~(\ref{eq:htotprime}).
\item Accept the new configuration with any detailed-balance satisfying rate, like the Metropolis rate.
\item Go to 1.
\end{enumerate}

The energy differences are given here for the four different cases.
\begin{itemize}
\item[(i)] $c_i=c_i'=1$, the particle diffuses within the surface layer.
\begin{equation}
\Delta \mathcal{H}_{\rm tot}' = \sum_{j \ne i}^M \left[ c_j \left( U_{\rm int}(|\vec{r}_i \; '-\vec{r}_j|) - U_{\rm int}(|\vec{r}_i
-\vec{r}_j|) \right) \right] + U_{\rm ext}(\vec{r}_i \; ') - U_{\rm ext}(\vec{r}_i) \;
\end{equation}
\item[(ii)] $c_i=c_i'=0$, the particle diffuses within the solution layer.
\begin{equation}
\Delta \mathcal{H}_{\rm tot}' = 0 \;
\end{equation}
\item[(iii)] $c_i=1$, $c_i'=0$, the particle desorbs.
\begin{equation}
\Delta \mathcal{H}_{\rm tot}' =
- U_{\rm ext}(\vec{r}_i) + \mu' -k_B T {\rm ln}{\left( M-N+1\right)}
-\sum_{j \ne i}^M c_j U_{\rm int}(|\vec{r}_i - \vec{r}_j|)
\end{equation}
\item[(iv)] $c_i=0$, $c_i'=1$, the particle adsorbs.
\begin{equation}
\Delta \mathcal{H}_{\rm tot}' =
U_{\rm ext}(\vec{r}_i) - \mu' +k_B T {\rm ln}{\left( M-N \right)}
+ \sum_{j \ne i}^M c_j U_{\rm int}(|\vec{r}_i - \vec{r}_j|)
\end{equation}
\end{itemize}
Here, $N$ represents the initial number of surface-layer particles,
corresponding to the configuration at step (1).

\clearpage
\begin{table}
\caption{
DFT calculations of the corrugation potential for Br/Ag(100) are shown in column~4.
More negative values indicate stronger binding,
and $(x,y)=(0,0)$ indicates the on-top site.
Points related by symmetry are not listed.
Column~5 lists the difference between the DFT calculations and the sinusoidal approximation ($\Delta=150$~meV)
of Eq.~(\ref{eq:corrpot}).
}
\vspace{0.2in}
\begin{tabular}{|l|l|l|l|l|}
\hline
$x/a$ & $y/a$ & site & $U_{\rm DFT}(x,y)$/meV & [$U_{\rm DFT}(x,y)-U(x,y)$]/meV \\ \hline
0 & 0 & on-top & 305 & 155 \\ \hline
0 & 0.25 &  & 206 & 131 \\ \hline
0 & 0.5 & bridge  & 0 & 0 \\ \hline
0.25 & 0.25 &  & 121 & 121 \\ \hline
0.25 & 0.5 &  & -54 & 21 \\ \hline
0.5 & 0.5 & hollow & -154 & -4 \\ \hline
\end{tabular}
\label{tab:dft}
\end{table}

\begin{table}
\caption{
Effective Arrhenius barriers for different corrugation amplitudes and damping parameters.
The errors are estimated from the linear fits shown in Fig.~\ref{fig:nubarr}.
}
\vspace{0.2in}
\begin{tabular}{|l|l|l|l|}
\hline
$\Delta$ [meV] & \multicolumn{3}{|l|}{Effective Barrier [meV]}\\ \cline{2-4}
 & $\alpha=1$ & $\alpha=10$ & $\alpha=100$ \\ \hline
100 & $85.2 \pm 0.8$ & $100 \pm 1$ & $98 \pm 2$ \\ \hline
150 & $125.0 \pm 0.7$ & $148 \pm 1$ & $144.9 \pm 0.8$ \\ \hline
200 & $164 \pm 1$ & $198 \pm 1$ & $199 \pm 1$ \\ \hline
\end{tabular}
\label{tab:effbarr}
\end{table}

\begin{table}
\caption{
Effective attempt frequency for different corrugation amplitudes and damping parameters.
The errors are estimated from the linear fits shown in Fig.~\ref{fig:nubarr}.
The time scale for one attempt is $\nu^{-1}$.
}
\vspace{0.2in}
\begin{tabular}{|l|l|l|l|}
\hline
$\Delta$ [meV] & \multicolumn{3}{|l|}{Effective $\nu$ [s$^{-1}$]}\\ \cline{2-4}
 & $\alpha=1$ & $\alpha=10$ & $\alpha=100$ \\ \hline
100 & $1.14 \pm 0.04 \times 10^{12}$ & $2.3 \pm 0.1 \times 10^{12}$ & $4.4 \pm 0.5 \times 10^{11}$ \\ \hline
150 & $1.31 \pm 0.04 \times 10^{12}$ & $2.9 \pm 0.2 \times 10^{12}$ & $5.8 \pm 0.2 \times 10^{11}$ \\ \hline
200 & $1.52 \pm 0.08 \times 10^{12}$ & $3.5 \pm 0.1 \times 10^{12}$ & $9.1 \pm 0.6 \times 10^{11}$ \\ \hline
\end{tabular}
\label{tab:effnu}
\end{table}


\clearpage

\begin{list}{}{\leftmargin 2cm \labelwidth 1.5cm \labelsep 0.5cm}

\item[\bf Fig. 1] The corrugation potential for Br adsorbed on Ag(100),
approximated as a two-dimensional sinusoidal function.
The grayscale indicates the Br binding energy,
with darker shades indicating more favorable binding and
lighter shades indicating less favorable binding.
The Ag surface atoms giving rise to the corrugation potential
are shown as circles with diameter $a=2.889$~{\AA}~\cite{OCKO97}.
Some of the circles have been removed for clarity.
Top, bridge, and hollow sites are labeled.

\item[\bf Fig. 2] Visualization showing both the solution layer and the surface layer.
The spheres indicate Br ions,
where the smaller inner sphere is shown for graphical purposes to indicate the center of the ion,
and the outer wireframe shell has radius $R_{\rm ion}=1.94$~{\AA}~\cite{CRSH90},
corresponding to the ionic radius of Br$^-$.
Br in the solution are confined to the upper plane and are non-interacting.
Br on the surface are confined to the lower plane and interact with
both the substrate corrugation potential and with other adsorbed Br particles.
The magnitude of the corrugation potential is shown by the grayscale
shading on the lower plane.  Darker shades indicate more favorable locations.

\item[\bf Fig. 3] Lateral interaction energy as a function of the lateral separation between a Br pair.
See Eq.~(\ref{eq:phi}).  The vertical lines indicate important distances as indicated in the figure.

\item[\bf Fig. 4] Root mean square (rms) displacement of Br adparticles from the four-fold-hollow sites vs coverage $\Theta$,
based on an off-lattice MC simulation in the canonical ensemble.
The system size is $L=32$, and the results for three different values of $\Delta$ are shown.
The displacement is largest for low coverages and small $\Delta$ and smallest for high coverages and large $\Delta$.

\item[\bf Fig. 5] Off-lattice grand-canonical MC isotherms for $L=32$ with three different values of $\Delta$.
The differences are most pronounced in the disordered region around $\Theta\approx 1/4$.

\item[\bf Fig. 6] Single-particle trajectory from a Langevin simulation with $\alpha=10$
for $\Delta=150$~meV at room temperature.
The gray-scale indicates the corrugation potential,
the heavy curve indicates a simulated particle trajectory,
and the dashed lines indicate the unit-cell boundaries used
in defining crossings.

\clearpage
\item[\bf Fig. 7] The crossing-time probability density from a Langevin simulation
with $\alpha=10$ and $\Delta=150$~meV for single-particle crossings at room temperature.
The linear region in the plot indicates the large-$\tau$ Arrhenius behavior.
Significant deviations from this behavior are seen at small $\tau$.

\item[\bf Fig. 8] Directional crossing probabilities from a Langevin simulation
with $\Delta=150$~meV and three different values of $\alpha$
for single-particle crossings at room temperature.
For small $\tau$,
the backward crossings indicate crossing/recrossing events,
and the forward crossings at small $\tau$ indicate running events.
The perpendicular direction includes both left and right crossings
and has been divided by two for comparison with the other crossing probabilities.
The figures show (a) $\alpha=1$, (b) $\alpha=10$, and (c) $\alpha=100$.

\item[\bf Fig. 9] Plot to determine the Arrhenius parameters from the Langevin single-particle crossing times.
The slope gives $E_{\rm barr}$, and the intercept gives $\nu$,
which are listed in Table~\ref{tab:effbarr} and Table~\ref{tab:effnu}, respectively.
The figures show (a) $\alpha=1$, (b) $\alpha=10$, and (c) $\alpha=100$.

\end{list}

\clearpage

\end{document}